\documentclass[letterpaper,preprint]{revtex4}
\usepackage{graphicx}
\usepackage{subfigure}
\begin{document}
\title{ 
Two-photon adiabatic passage in ultracold Rb interacting with a single nanosecond, chirped pulse}

\author{Gengyuan Liu, S.A. Malinovskaya\\ Department of Physics and Engineering Physics, Stevens Institute of Technology, Hoboken, New Jersey 07030, USA}

\begin{abstract}
A semiclassical, four-level model of a nanosecond, chirped pulse interacting with all optically accessible hyperfine states in the ultracold Rb atom is analyzed aiming at population inversion within $5S_{1/2}$ electronic state. The nature of two-photon adiabatic passage performed by such a single pulse having  a bandwidth smaller than the hyperfine splitting of $5S_{1/2}$ state is investigated in the framework of the dressed state picture. It is shown that two dressed states are involved in the adiabatic dynamics of population inversion. The excited state manifold appeared to play an important mediating role in the  mechanism of population transfer. 
\end{abstract}
\maketitle

Adiabatic passage known to be a vital control technique for the manipulation of dynamics in atoms and molecules and the preparation of  predetermined superposition states in these systems. It is broadly implemented in control schemes at ultracold temperatures. 
Examples include the use of the pulse area solution for fast two-qubit phase gates \cite{Ja00}, a creation of ultracold molecules via stimulated Raman adiabatic passage (STIRAP) \cite{Ni08}, and the implementation of different combinations of chirped pulses to deterministically excite Rydberg states \cite{Mo08,Be11,Ku14}. 
Here we discuss control of valence electron in ultracold Rb 
by a single nanosecond pulse whose bandwidth is much narrower than the transitional frequency between hyperfine levels of $5^{2}S_{1/2}$ state. The population inversion is achieved through the Raman transitions that involve hyperfine structure of $5^{2}P_{1/2}$ or $5^{2}P_{3/2}$ states and with the aid of linear chirping of the pulse. The adiabatic solution is found through a systematic numerical analysis of the response of a four-level system, representing all  optically attainable hyperfine states, on a broad variation of the field parameters. 
We have made a detailed analysis of the dressed state picture to gain insight into adiabatic mechanism of the two-photon Raman transition by means of a single narrow-band nanosecond pulse, which reveals the involvement of two dressed states into adiabatic passage producing population inversion. These two dressed states form a subset owing to two energetically close hyperfine states of the transitional $5^{2}P_{1/2}$ or $5^{2}P_{3/2}$ state. Because only one chirped pulse is implemented having the bandwidth narrower than the hyperfine splitting of $5^{2}S_{1/2}$ state, the excited state manifold plays the key role in the passage as a mediator, thus, distinguishing this approach from previous experiences. 

\begin{figure}
\centerline{\includegraphics[width=9cm]{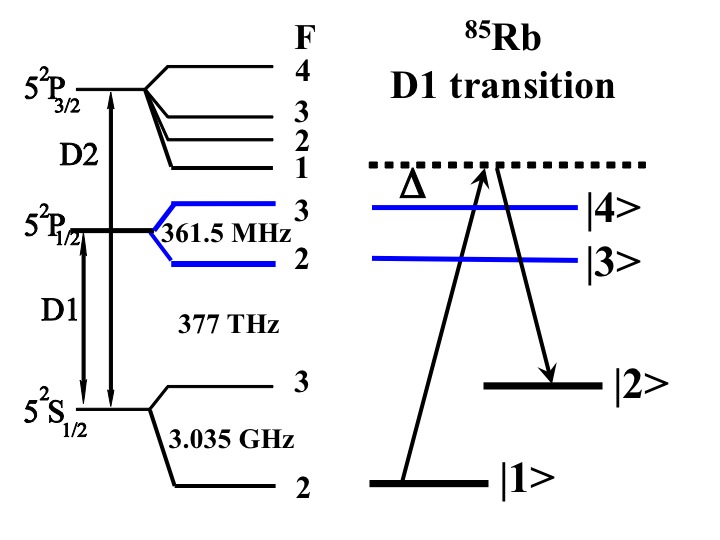}}
\caption{ Four optically attainable hyperfine states of $5S$ and $5P$ shells with the energy differences that correspond to the D1 line. Initially, the population is in the ground state $\left|1\right\rangle $. Note that the hyperfine splitting of the $5S_{1/2}$ orbital is approximately an order of magnitude greater than the splitting of $5P_{1/2}$ orbital, \cite{key-1}. }\label{4-level system}
\end{figure}

We consider the four-level system which takes into account all optically allowed transitions between hyperfine states belonging to $5^{2}S_{1/2}$ and $5^{2}P_{1/2}$ or $5^{2}P_{3/2}$ states, Fig.(\ref{4-level system}). The Hamiltonian that describes the four-level system interaction with a single chirped nanosecond pulse reads

\begin{equation}
\hat{H_{int}}=h \left[\begin{array}{cccc}
\Delta+\omega_{43}+\alpha(t-T) & 0 & -\Omega_{R}(t)/2 & -\Omega_{R}(t)/2\\
0 & \Delta+\omega_{43}+\omega_{21}+\alpha(t-T) & -\Omega_{R}(t)/2 & -\Omega_{R}(t)/2\\
-\Omega_{R}(t)/2 & -\Omega_{R}(t)/2 & 0 & 0\\
-\Omega_{R}(t)/2 & -\Omega_{R}(t)/2 & 0 & \omega_{43}\end{array}\right]
\label{4lvlHam}
\end{equation}.

\begin{figure}
\centerline{\includegraphics[width=9cm]{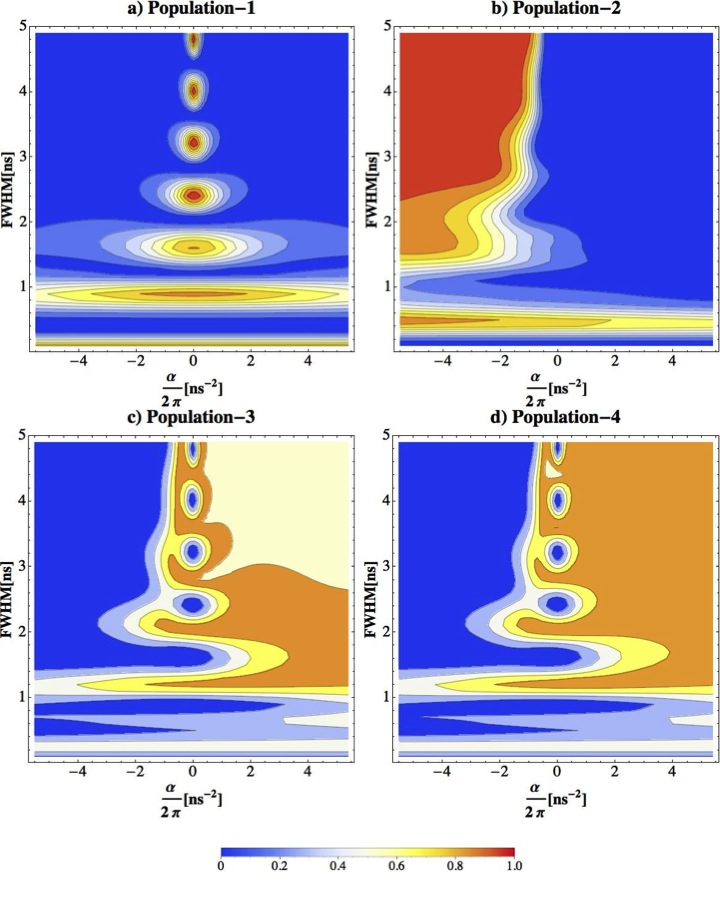}}
\caption{ The end-of-pulse population distribution in the four-level system,
achieved via two-photon transitions using a single, linearly chirped laser pulse. The values of the system parameters
are  $\omega_{21}$ = 3.035 GHz, $\omega_{43}$ = 0.362 GHz, characteristic for $^{85}$Rb \cite{key-1}, the peak Rabi frequency is $\Omega_R$=3.035 GHz, and one-photon detuning $\Delta$ is zero.  } \label{contour}
\end{figure}

Here $\Omega_{R}(t)\equiv - \mu E_{0}(t)/h$ is the Rabi frequency with the peak value $\Omega_R$, $\alpha$ is the linear chirp parameter, $\alpha/2\pi$ has units Hz/s, and $\Delta$ is the one-photon detuning. Solving numerically the time-dependent Schr\"{o}dinger equation 
with the Hamiltonian in Eq.(\ref{4lvlHam}) for various values of the field parameters provides an accurate picture of light-matter interactions and allows for finding the exact values of the field parameters required to obtain a predetermined non-equilibrium superposition state, population invention or population return. It also reveals the adiabatic region of population transfer to the target state $|2>$, which is the upper hyperfine state F=2 of the $5^{2}S_{1/2}$. Populations of the four states at the end of the pulse as a function of the pulse chirp rate and the full width at the half maximum (FWHM) are presented in Fig.(\ref{contour}), \cite{Li14}.  FWHM of the Gaussian pulse relates to the pulse duration $\tau_0$ as FWHM=$\tau_0 2\sqrt{ \ln2}$, $\omega_0$. The adiabatic region of light-matter interaction leading to population inversion is observed for parameters that satisfy the adiabaticity conditions $ |\alpha/(2\pi) | \tau_0 > \omega_{21}$ and $|\alpha / (2 \pi)| < \Omega_R^2$. The physical values applicable to $^{85}$Rb are, e.g., the peak Rabi frequency  $\Omega_R = \omega_{21}$ (3.035 GHz), the chirp rate $\alpha/2\pi$=-0.3[$\omega_{21}^2$] (-3 GHz/ns) or faster and the pulse duration $\tau_0 \geq 5.5 \omega_{21}^{-1}$, ($\geq$ 1.8 ns). The negative value of the chirp rate is well understood since we start from the one-photon  blue detuning with the largest transition frequency $\omega_{41}$ and gradually decrease the instantaneous frequency to pass through each one-photon resonance, first with $\omega_{41}$, then with $\omega_{31}$, $\omega_{42}$ and $\omega_{32}$. The exemplified field parameters may be obtained in modern experimental setups such as described in, e.g., \cite{Ro07}. Since the spectral bandwidth of the nanosecond pulse (~0.5 GHz for 1.8 ns pulse) is much narrower than the energy separation between the hyperfine states (3.035 GHz) of the $5^{2}S_{1/2}$, a question of fundamental interest arises as to what is the mechanism of the adiabatic population transfer performed with the two photons that are never present in the system with the frequency "right" to satisfy the two-photon resonant condition? We performed the dressed state analysis to gain insight into the adiabatic and nonadiabatic nature of quantum control of population dynamics in the four-level system using a single narrowband but chirped laser pulse. 

We first outline a basic concept of the dressed state analysis, \cite{Be112,Al08}, and its extension to the case when  adiabatic passage may occur within a subset of dressed states coupled to each other. 
A wave function of a quantum system $|\Psi(t)>$ may be written as a linear superposition of the bare states in the field interaction representation $|i>$ with the respective probability amplitudes $C_i$  

\begin{equation}
|\Psi(t)> = \Sigma_i^4 C_i |i>.
\end{equation}

The time-dependent Schr\"{o}dinger equation then reads as $i \hbar {\bf {\dot{C}} }= \hat{H}_{int} {\bf C}$.  We apply a unitary transformation   ${\bf T}$ 
 to the $\hat{H}_{int}$ leading to diagonalization of the Hamiltonian. Here, ${\bf T}$ is an eigenvector matrix of $\hat{H}_{int}$. The obtained Hamiltonian is the so called dressed state Hamiltonian  $\hat{H_d} = {\bf T^{\dag} } \hat{H}_{int} {\bf T}$ written in the basis of the dressed states $|I>$ such that $|\Psi(t)> = \Sigma_i^4 C_{di} |I>$ and $ {\bf C_{d} = T C}$. Then, putting the reverse expression  ${\mathbf{C}}=T^{+}\mathbf{C}_d $ into the Schr\"{o}dinger equation and assuming that all quantities are time dependent we arrive at

\begin{equation}
i\hbar \dot{\mathbf{C}}_d  = \hat{H_d} \mathbf{C}_d  - i\hbar \hat{T}\dot{\hat{T}}^{+}  \mathbf{C}_d.
\label{eq:7}
\end{equation}
Since the Hamiltonian $\hat{H_d}$ is diagonal, the dressed states would evolve without mixing with each other if to  disregard the second term on the right side. This is the essence of the adiabatic approximation, when the system, once placed in a selected dressed state by the initial conditions, continues evolution within this dressed state only. The second term is responsible for the nonadiabatic coupling between the dressed states, it contains matrix operator $\hat{T}\dot{\hat{T}}^{+}$ which is non-diagonal and determines the degree of non-adiabatic mixture between the dressed states.  If the matrix elements of the $\hat{T}\dot{\hat{T}}^{+}$ are much less than the energy splittings between the respective dressed states, the dynamics may be considered as adiabatic. 
Analises of the time-dependence of the dressed state energies and the wave functions as well as a comparison of non-adiabatic and adiabatic terms help to estimate the degree of adiabaticity and a possibility for quantum control. From another hand, if to aim to find the field parameters that provide the adiabatic solution, it is useful to move to the dressed state basis and within the adiabatic approximation find the field conditions and parameters for dynamics in a single dressed state. When implemented in the exact Schr\"{o}dinger equation, these parameters may yield quasi-adiabatic behavior within the exact Schr\"{o}dinger picture. 

In a multi-level case, the non-adiabatic coupling may be small for some dressed states, but significant for the others. Then coupled dressed states may create a subsystem within which the dynamics occurs adiabatically. For the four-level system described by the Hamiltonian in Eq.(\ref{4lvlHam}),  two  dressed states, $|I>$ and $|III>$, are coupled in the vicinity of the avoided crossing and provide adiabatic passage, while two other dressed states stay intact: 
\begin{equation}
\label{HAM34} 
i\hbar 
\left( {\begin{array}{c}
   {\dot{C_{dI}} }  \\
   {\dot{C_{dII}}}  \\  {\dot{C_{dIII}}}  \\  {\dot{C_{dIV}}}  \end{array}}
\right) 
 = \left( \begin{array}{cccc} \lambda_1& 0 & V & 0\\
0 & \lambda_2 & 0 & 0\\ V & 0 & \lambda_3 & 0\\
0 & 0 & 0 & \lambda_4\\
\end{array} \right) 
\left( {\begin{array}{c}
   {C_{dI} }  \\
   {C_{dII}}  \\  {C_{dIII}}  \\  {C_{dIV}}  \end{array}}
\right) 
\end{equation}

Note, that a superposition of two active dressed states, $|I>$ and $|III>$, may be approximated by a single dressed state if to reduce the four-level system to a three-level $\Lambda$ system by substituting two energetically close excited states by a single one. We will discuss this approximation in details below.  In the framework of the three-level $\Lambda$ system, the adiabatic dynamics takes place within a single dressed state having the energy dependence on time resembling that of a superposition 
state in the four-level system. 

We demonstrate the concept by analyzing the time dependence of the dressed state energies in the four-level system and the squares of the respective eigenvector elements $T_{ij}$ that show the time evolution of the population of the bare states within each dressed state. The parameters of the nanosecond chirped pulse that provide the adiabatic passage are chosen from the numerical solution of the Schr\"{o}dinger equation, Fig.(\ref{contour}). 
Here a broad adiabatic region leading to population inversion to the upper hyperfine state of the $5S_{1/2}$ state is observed for the parameters starting from FWHM = 2.5 ns and higher and the absolute value of the chirp rate  $|\alpha/2\pi|$ grater than 2 GHz/ns. 

\begin{figure}
\centerline{
\includegraphics[width=10cm]{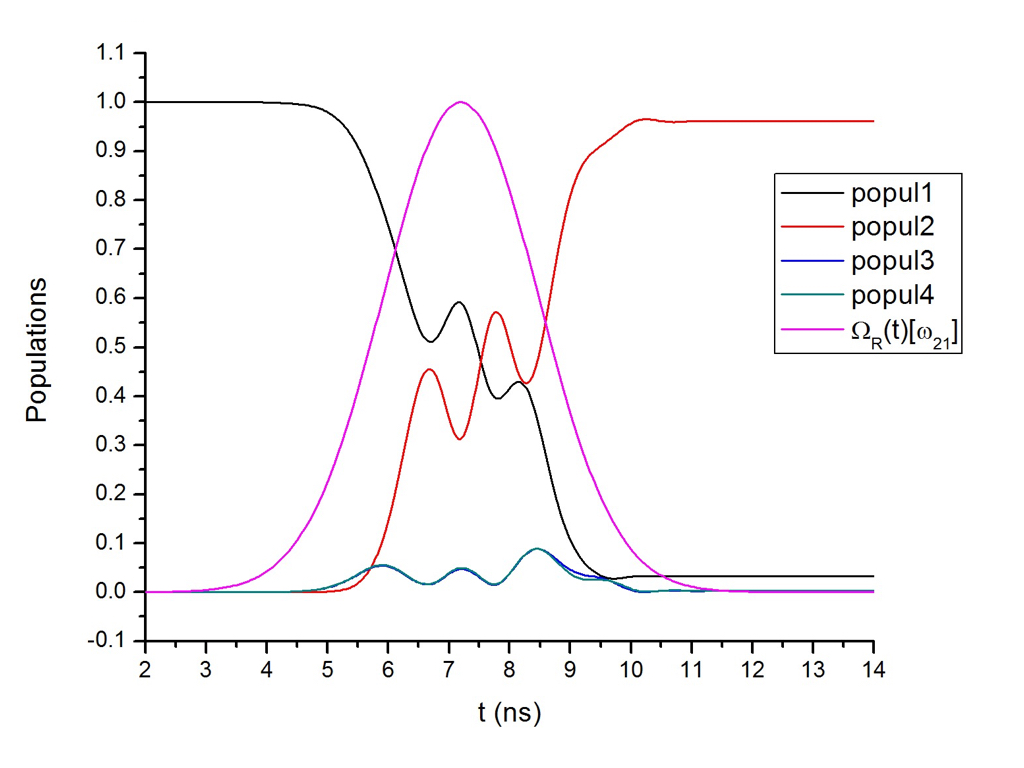}}
\caption{ Time-dependent picture of the population dynamics in four hyperfine states demonstrating adiabatic passage to the final state $|2>$. The field parameters are $\Omega_R$=3.035 GHz, FWHM=2.995 ns and $\alpha/(2\pi)$=-2.947 GHz/ns.}    \label{adiabTD}
\end{figure}

The exact time-dependent picture of the adiabatic passage to the final hyperfine sate obtained by numerically solving the time-dependent Schr\"{o}dinger equation with the Hamiltonian in Eq.(\ref{4lvlHam}) is shown in Fig.(\ref{adiabTD}) for parameters from the adiabatic region, e.g., FWHM= 2.995 ns, $\alpha/2\pi$ = -2.947 GHz/ns, $\Omega_R$=3.035 GHz and $\Delta=0$. Population flow from the ground $|1>$ to the excited $|2>$ state begins at about half-way before the Rabi frequency reaches the peak value.  It is owing to the one-photon off-resonance contribution of light-matter interaction  into the population passage. Notably, population inversion dynamics follows the excitation of states $|3>$ and $|4>$; those states get transitionally populated to up to 10$\%$ and show oscillations. The oscillations are attributed to small nonadiabatic coupling between all dressed states. They vanish with the increase of the Rabi frequency. 

\begin{figure}
\centerline{
\includegraphics[width=10cm]{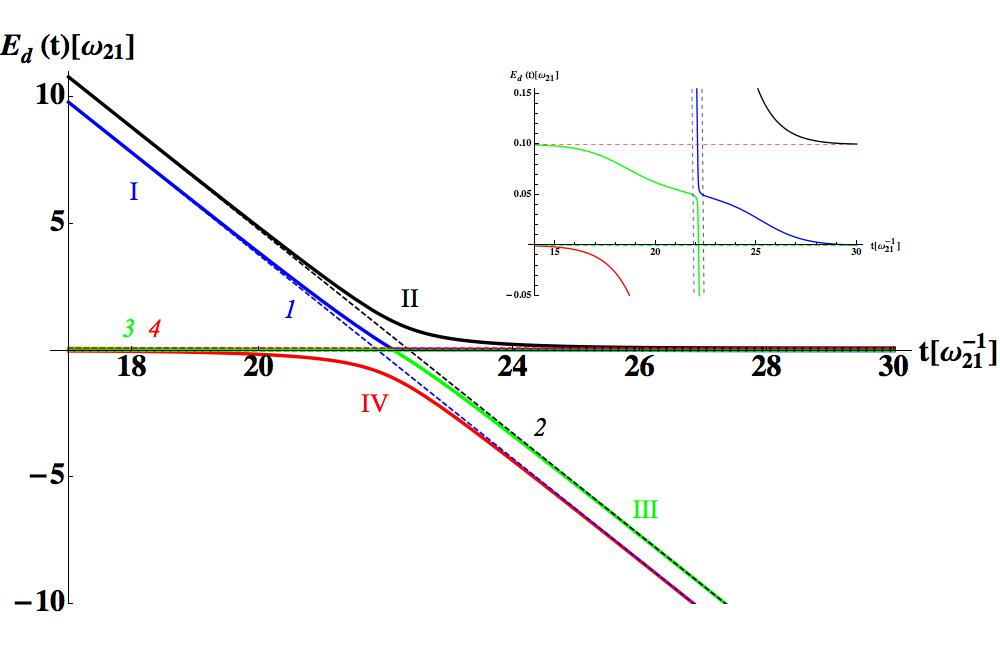}}
\caption{ Dressed state energies as a function of time that lead to adiabatic passage to the final state $|2>$. The field parameters are $\Omega_R$=3.035 GHz, FWHM=2.995 ns and $\alpha/(2\pi)$=-2.947 GHz/ns.}    \label{adiabDSE}
\end{figure}

The respective dressed state energies  for the same field parameters FWHM= 2.995 ns, $\alpha/2\pi$ = -2.947 GHz/ns, $\Omega_R$=3.035 GHz and $\Delta=0$ are depicted in Fig.(\ref{adiabDSE}).
It demonstrates that if the chirp rate is large enough, the change of the instantaneous carrier frequency is sufficient to swipe adiabatically through the two-photon resonance. The dynamics in the system begins within the blue dressed state $|I>$, which coincides with initially populated bare state $|1>$, (shown in blue dashed color). Further on, the blue dressed state $|I>$ approaches  the green dressed state $|III>$ near the peak value of the field amplitude to form an avoided crossing.  In its vicinity, population moves efficiently from the blue $|I>$ to the green $|III>$ dressed state and, thus, resides transitionally on the excited bare state manifold for a restricted period of time before moving to the final bare state $|2>$  as the green dressed state $|III>$ evolves to become $100 \%$ constituted of it. Thus, the excited states $|3>$ and $|4>$ keep population for the time needed for the instantaneous frequency to acquire the value needed to accomplish the two-photon resonance. In such a way, the adiabatic passage is performed by a subset of coupled through the avoided crossing dressed states and requires the excited state manifold to mediate the dynamics owing to only one chirped, narrowband pulse used to perform the inversion.



\begin{figure}
\centerline{
\includegraphics[width=10cm]{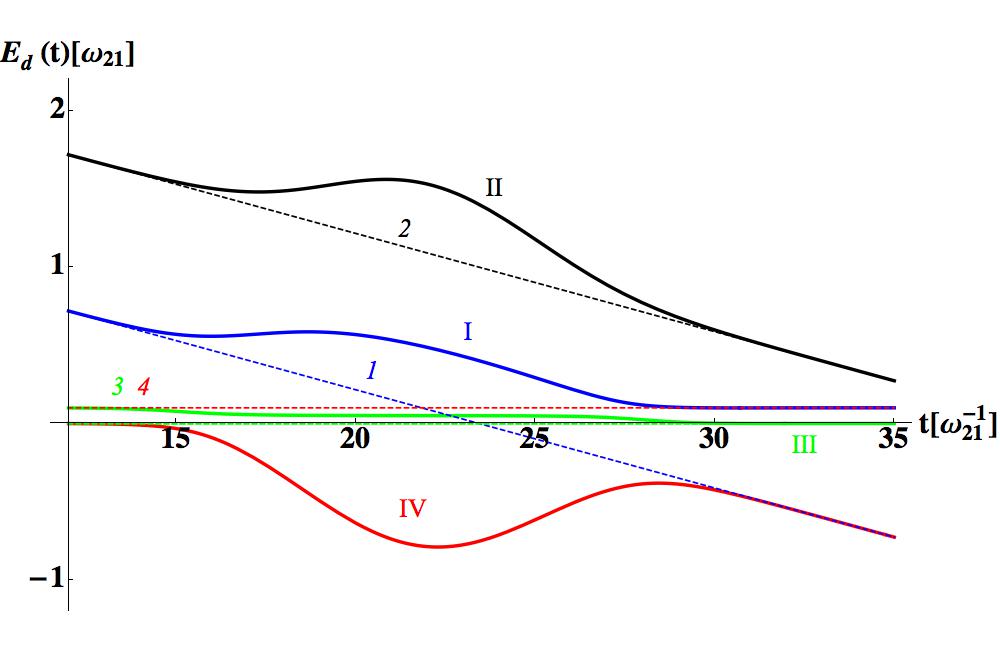}}
\caption{ Dressed state energies as a function of time for the field parameters $\Omega_R$ = 3.035 GHz, FWHM = 2.995 ns and $\alpha/(2\pi)$ = -0.092 GHz/ns. No adiabatic passage occurs to the final bare state $|2>$ and the dynamics is preserved within dressed state $|I>$, which coincides with bare states $|3>$ and $|4>$ at the end of the pulse. }    \label{nonadiabDSE}
\end{figure}

For a comparison, the case when the chirp is not large enough to provide the range of frequencies needed to satisfy the two-photon transition is shown in Fig(\ref{nonadiabDSE}). Here the chirp rate is $\alpha/2\pi$ = -0.092 GHz/ns and the pulse duration is $\tau_0$ = 1.799 ns (FWHM = 2.995 ns), giving $\alpha \tau_0 $= 0.166 GHz $ \ll \omega_{21}$ = 3.035 GHz.  The other parameters are the peak Rabi frequency $\Omega_R$ = $\omega_{21}$=3.035 GHz,  and one-photon detuning $\Delta$ = 0. Here, blue dressed state initially coincides with blue bare state $|1>$; this picture remains for most of the pulse duration. Then, close to the exponential end of the pulse amplitude, blue dressed state approaches the green dressed state, which is mainly a superposition of the excited states $|3>$ and $|4>$, and nonadiabatically transfers a fraction of population to them before the pulse ceases. 
As the result, the population remains mostly in the initial, ground state owing to the field parameters not satisfying the adiabaticity condition for population inversion determined by the condition $\alpha/(2\pi) \tau_0 > \omega_{21}$, 
even though the Landau-Zener condition is satisfied, $\Omega_R^2 / |\alpha| \sim 3/2 \times 10^2$. 

\begin{figure}
\centerline{\includegraphics[width=9cm]{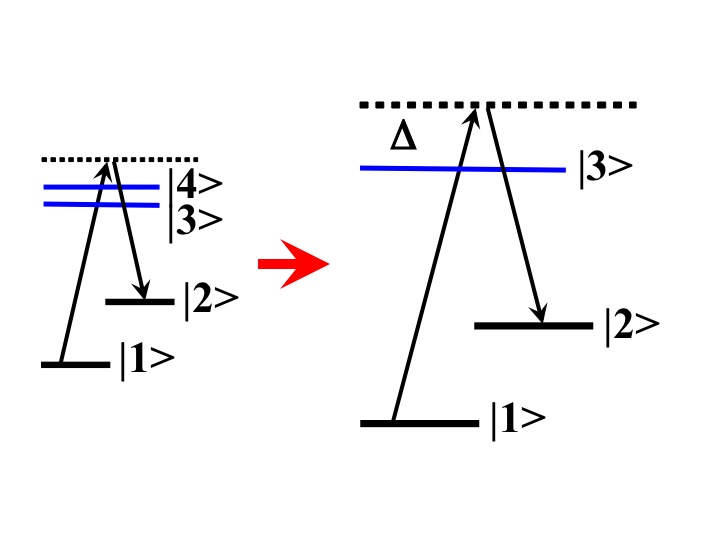}}
\caption{ A three-level $\Lambda$ system considered as an approximate model with two optically attainable hyperfine states of $5P$ shell described by a single transitional state $|3>$. The approximation is implemented owing to the hyperfine splitting of the $5P_{1/2}$ orbital being about an order of magnitude less than the splitting of $5S_{1/2}$ orbital. It is valid under the conditions that $ |\alpha/(2\pi) | \tau_0 \gg \omega_{43}$ and  $\Omega_R \gg \omega_{43}$. Initially, the population resides in the ground state $\left|1\right\rangle $.  }\label{3lvl-scheme}
\end{figure}

To benefit from an analytical solution, we performed a comparative analysis with an effective three-level $\Lambda$ system, Fig.(\ref{3lvl-scheme}) that works as a good approximation to the four-level system giving a qualitatively similar quantum yield when the pulse duration and the chirp rate satisfy $ |\alpha/(2\pi) | \tau_0 \gg \omega_{43}$ and  $\Omega_R \gg \omega_{43}$. A detailed numerical solution for the three-level $\Lambda$ system is discussed in \cite{Co12,Co13}. 

From the Hamiltonian in Eq.(\ref{4lvlHam}), we may easily get the following set of equations for the probability amplitudes in the field interaction representation assuming the one-photon detuning $\Delta$=0
\begin{eqnarray}\label{(eq3)}
&i\dot{a}_1=\alpha(t-T)a_1-\Omega_R(t)(a_3+a_4)/2 \nonumber \\
&i\dot{a}_2=[\omega_{21}+\alpha(t-T)]a_2-\Omega_R(t)(a_3+a_4)/2 \nonumber \\
&i\dot{a}_3=-\Omega_R(t)(a_1+a_2)/2-\omega_{43}a_3\\
&i\dot{a}_4=-\Omega_R(t)(a_1+a_2)/2 \nonumber
\end{eqnarray}
 
By making a substitution

\begin{eqnarray}
&(a_3+a_4)/\sqrt{2}=a_+\\
&(a_3-a_4)/\sqrt{2}=a_- ,
\end{eqnarray}

we arrive at the following relation for $a_+$ and $a_-$
\begin{eqnarray}
&i(\dot{a}_+)=i(\dot{a}_3+\dot{a}_4)/\sqrt{2}=-\Omega_R(t)(a_1+a_2)/\sqrt{2}-\omega_{43}/\sqrt{2}a_3\\
&i(\dot{a}_-)=i(\dot{a}_3-\dot{a}_4)/\sqrt{2}=-\omega_{43}/\sqrt{2}a_3.
\end{eqnarray}

If $\omega_{43}\sqrt{2}$ is small enough compared to $\Omega_R$, it may be neglected. Then, $a_3-a_4=const$, and $a_-$ may be omitted from the dynamics calculation. Finally, the Eqs. $(\ref{(eq3)})$ are reduced to a set of three coupled differential equations

\begin{eqnarray}\label{(eq6)}
&i\dot{a}_1=\alpha(t-T)a_1-\Omega_R(t)/\sqrt{2} a_+ \\
&i\dot{a}_2=[\omega_{21}+\alpha(t-T)]a_2-\Omega_R(t)/\sqrt{2} a_+\\
&i\dot{a}_+=-\Omega_R(t)/\sqrt{2} (a_1+a_2)
\end{eqnarray}

The Hamiltonian for the three-level approximation in the field interaction representation reads

\begin{equation} \label{eq7}
\hat{H}_{int}=h \left[\begin{array}{cccc}
\alpha(t-T) & 0 & -\Omega_{R}(t)/\sqrt{2}\\
0 & \omega_{21}+\alpha(t-T) &  -\Omega_{R}(t)/\sqrt{2}  \\
 -\Omega_{R}(t)/\sqrt{2} &  -\Omega_{R}(t)/\sqrt{2} & 0 
\end{array}\right]
\end{equation}

\begin{figure}
\centerline{\includegraphics[width=5in]{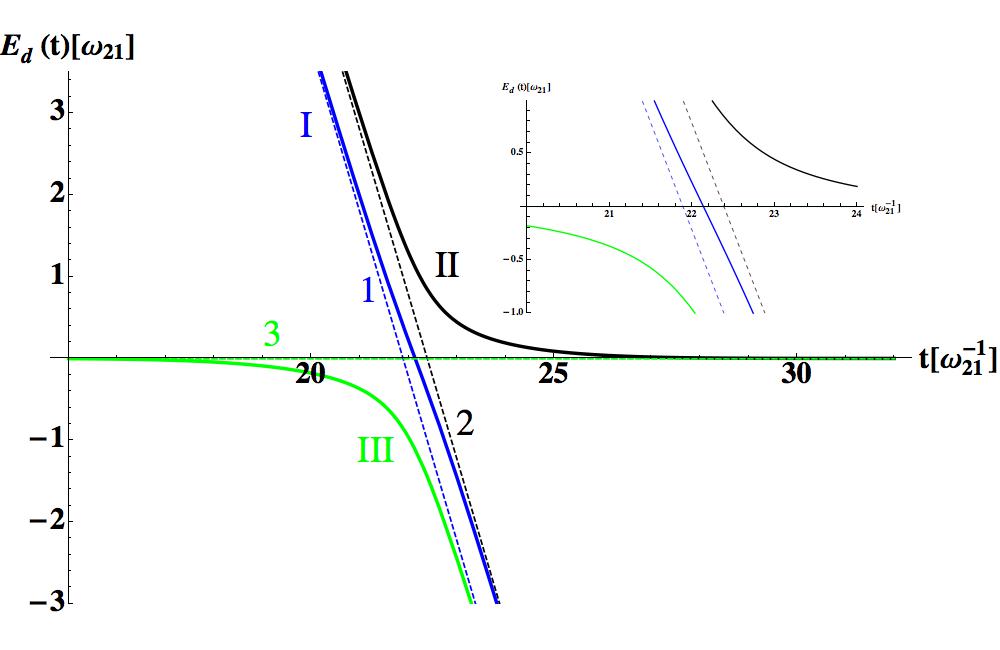}}
\caption{The adiabatic dynamics of population transfer between the bare states within each dressed state in the three-level $\Lambda$ system. The field parameters are $\Omega_R$=3.035 GHz, FWHM = 2.995 ns and $\alpha/(2\pi)$ = -2.947 GHz/ns.   }\label{adia3a}
\end{figure}

\begin{figure}
\centerline{\includegraphics[width=5in]{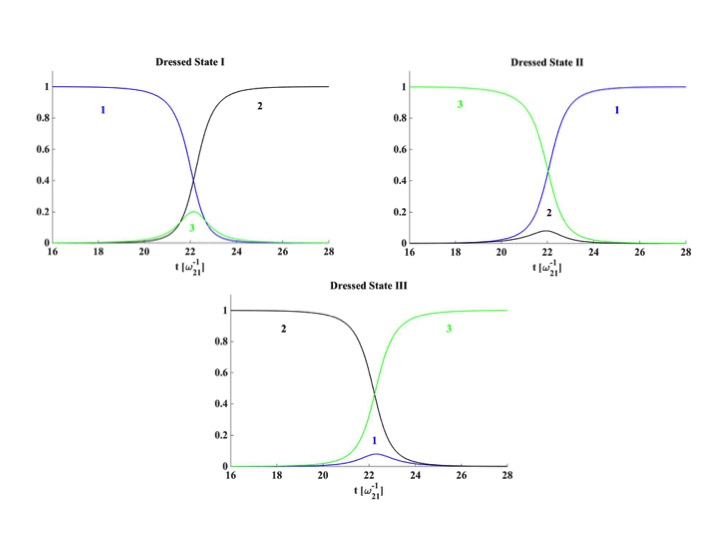}}
\caption{The adiabatic dynamics of population transfer between the bare states within each dressed state. Population inversion between states $|1>$ and $|2>$ is observed within the dressed state $|I>$, (a). The field parameters are $\Omega_R$ = 3.035 GHz, FWHM = 2.995 ns and $\alpha/(2\pi)$ = -2.947 GHz/ns.   }\label{ad3pa}
\end{figure}

Analytical diagonalization of the Hamiltonian in Eq.(\ref{eq7}) leading to expressions for the dressed state energies and respective eigenfunctions showed no dark state solution as it was the case, e.g., in conventional STIRAP scheme, \cite{Kr07}. The time-dependent wave function describing each dressed state contains nonzero probability amplitudes for all three bare states. Since the expressions for the dressed state energies and the probability amplitudes look heavy, we do not present them here, but rather show their time dependence obtained numerically.  A numerical analysis of the dressed states was performed for the three-level $\Lambda$ system described by the Hamiltonian in Eq.(\ref{eq7}) within the same range of parameters as for the four-level system. As an example, the results for $\alpha/2\pi$ = -2.947 GHz/ns, $FWHM$ = 2.995 ns and $\Omega_R$ = 3.035 GHz are discussed in more details. The time-dependence of the dressed state energies is depicted in Fig.(\ref{adia3a}).  Here the dynamics occurs within a single dressed state $|I>$, shown in blue color. Initially, it coincides with bare state $|1> $ (dashed blue), followed by adiabatic transition within the same dressed state in the vicinity of the peak values of the pulse amplitude from the bare state $|1> $ through the excited state $|3>$ (dashed green) to final bare state $|2>$ (dashed black).  In this approximate three-level model, the dressed state $|I>$ plays the role of the subset of two dressed states, $|I>$ and $|III>$, in the four-level case. In Fig.(\ref{ad3pa}), the adiabatic dynamics of population transfer between the bare states within each dressed state is shown by the time dependence of  $T_{ij}^2$. It reveals a smooth adiabatic passage from the initial bare state $|1> $ to the final bare state $|2> $ in the dressed state $|I>, $Fig.(\ref{ad3pa}a). 

\begin{figure}
\centerline{\includegraphics[width=5in]{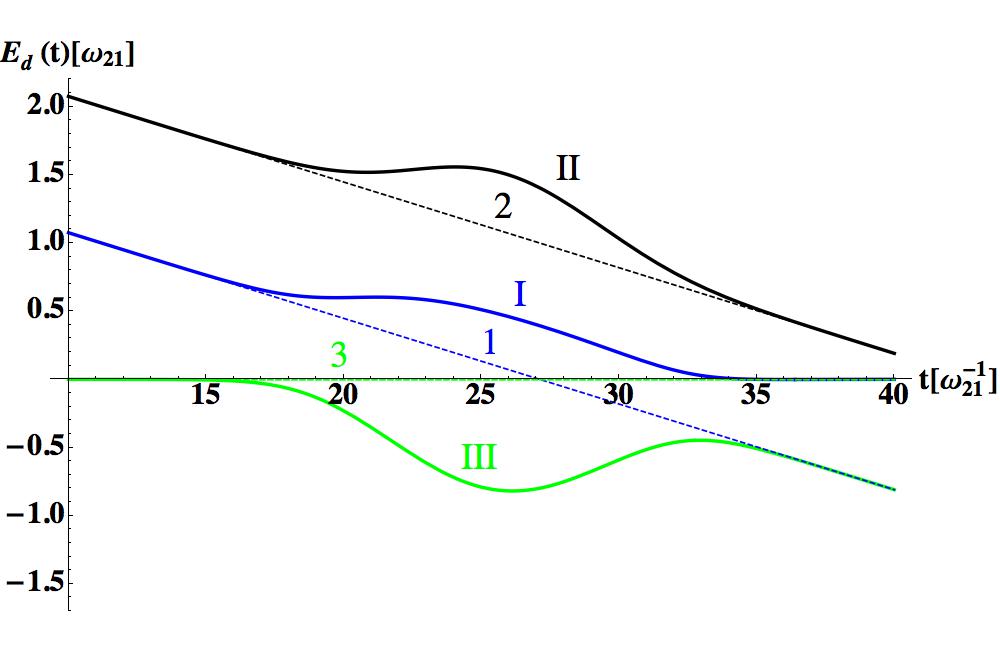}}
\caption{Dressed state analysis of energy picture for the three-level. The field parameters are $\Omega_R$ = 3.035 GHz, FWHM = 2.995 ns and $\alpha/(2\pi)$ = -0.092 GHz/ns.   }\label{nadia3a}
\end{figure}

\begin{figure}
\centerline{\includegraphics[width=5in]{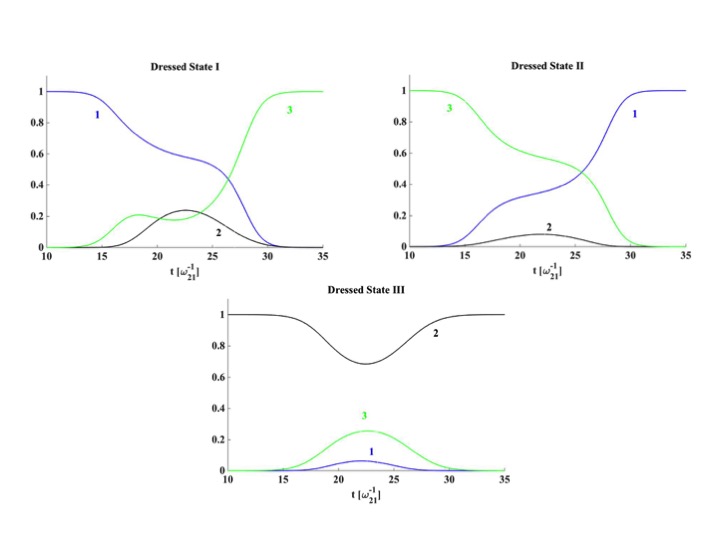}}
\caption{The adiabatic dynamics of population transfer between the bare states within each dressed state which does not lead to population inversion between states $|1>$ and $|2>$.  The field parameters are $\Omega_R$ = 3.035 GHz, FWHM = 2.995 ns and $\alpha/(2\pi)$ = -0.092 GHz/ns.   }\label{nad3pa}
\end{figure}

However, if to choose a set of parameters such that they do not satisfy the condition $\alpha/(2\pi) \tau_0 > \omega_{21}$, which is, for example, $\alpha/2\pi$ = -0.092 GHz/ns, FWHM = 2.995 ns, the dynamics still occurs within a single dressed state but does not lead to the population inversion to the final bare state $|2>$. In the beginning, the energy of dressed state $|I>$ coincides with the bare state $|1>$; however, the pulse ceases before the dynamics within dressed state $|I>$ progresses to the final bare state $|2>$, which makes it retaining in the intermediate state $|3>$, Fig.(\ref{nadia3a}). The chirp rate of the pulse is not fast enough to switch through the two-photon resonance. The bare state population dynamics within each of three dressed states is shown in Fig.(\ref{nad3pa}), it supports this outcome by manifesting adiabatic passage from the initial bare state $|1>$ to the intermediate bare state $|3>$ within the active dressed state $|I>$, Fig.(\ref{nad3pa}a).

In summary, we have analyzed the mechanism of population dynamics in a new method for population inversion within the hyperfine structure in alkali atoms at ultracold temperatures that relies on implementing a single nanosecond chirped pulse of intensity on the order of kW/cm$^2$ having the bandwidth much narrower than the two-photon transition frequency in the atomic system. The results are based on the  developed semiclassical model of the pulse interaction with the four-level system representing all optically accessible   hyperfine states of $5^{2}S_{1/2}$ and $5^{2}P_{1/2}$ or $5^{2}P_{3/2}$ states in ultracold $^{85}$Rb. The adiabatic passage leading to population inversion is achieved for parameters that satisfy the condition $ |\alpha/(2\pi) | \tau_0 > \omega_{21}$ and the Landau-Zener adiabaticity condition $|\alpha / (2 \pi)| < \Omega_R^2$. Dressed state analysis was performed to gain understanding about the mechanisms of two-photon Raman transitions performed by a single, narrowband, chirped pulse having the bandwidth $\Delta \omega \ll \omega_{21}$. It revealed an existence of a subset of dressed states coupled in the vicinity of avoided crossings that perform the adiabatic passage leading to the population inversion. 
When considered as an approximation of the four-level system, the three-level $\Lambda$ system demonstrates population inversion  within a single dressed state. This dressed state resembles the time-dependence of two dressed states in the active subset in the four-level system. This justifies the validity of several dressed states approach to adiabatic passage.   

The authors acknowledge fruitful discussions with Vladimir Malinovsky, Elena Kuznetsova and Phillip Gould. This research was supported by the National Science Foundation under Grants No. NSF-KITP-14-224 and No. NSF PHY12-05454.

\end{document}